\documentclass[fleqn,usenatbib]{mnras}
\usepackage{newtxtext,newtxmath}

\usepackage[T1]{fontenc}
\usepackage{graphicx}	
\usepackage{amsmath}	
\usepackage{amssymb}	

\usepackage{times}
\usepackage{epsfig}
\usepackage{psfrag}
\usepackage{ulem}
\usepackage{tikz}
\usepackage{etaremune}
\usepackage{pifont}
\usepackage{colortbl}
\usepackage{soul}

\usepackage{hyperref}

\title[3D numerical simulations of structured GRB jets]{Three-dimensional numerical simulations of structured GRB jets}

\author[Urrutia, De Colle \& L\'opez-C\'amara]{
Gerardo Urrutia$^{1}$\thanks{E-mail: gerardo.urrutia@nucleares.unam.mx},
Fabio De Colle$^{1}$, and
Diego L{\'o}pez-C{\'a}mara$^{2}$
\\
$^{1}$Instituto de Ciencias Nucleares, Universidad Nacional Aut{\'o}noma de M{\'e}xico, A. P. 70-543 04510 D.F. Mexico\\
$^{2}$C\'atedras CONACyT -- Universidad Nacional Aut\'onoma de M\'exico, Instituto de Astronom\'ia, AP 70-264, CDMX  04510, M\'exico\\
}
\date{Accepted XXX. Received YYY; in original form ZZZ}

\pubyear{2022}

\begin{document}
\label{firstpage}
\pagerange{\pageref{firstpage}--\pageref{lastpage}}
\maketitle

\begin{abstract}
After the detection of GRB 170817A, the first unambiguous off-axis gamma-ray burst (GRB), several studies tried to understand the structure of GRB jets. The initial jet structure (directly produced by the central engine) can be partially preserved, or can be completely modified by the interaction with the environment. 
In this study, we perform three-dimensional, special relativistic hydrodynamics simulations of long GRB jets evolving through a massive progenitor star. Different jet scenarios were considered: Top-hat, Gaussian jets dominated by pressure or by kinetic energy, as well as a model of a supernova (SN) plus a jet both propagating through the progenitor.
We found that, while propagating inside the progenitor star, jets with different initial structures are nearly indistinguishable. Kinetic dominated jets are faster and more collimated than pressure dominated jets. The dynamics of jets inside the progenitor star strongly depends on the presence of an associated SN, which can substantially decelerate the jet propagation. We show that the initial structure of GRB jets is preserved, or not, mainly depending on the jet collimation. The initial structure is preserved in uncollimated jets, i.e. jets which move through low density environments. Meanwhile, jets which move through dense environments are shaped by the interaction with the medium and remain collimated.

\end{abstract}

\maketitle

\section{Introduction}
\label{sec:intro}
Gamma-Ray Bursts (GRBs) are bright, high-energy transients emitted by highly relativistic jets \citep[see, e.g.,][]{kumar15, Levan2018}. GRBs are classified as short and long (SGRBs and LGRBs, respectively) based on their duration. SGRBs are associated with neutron star mergers and, possibly, neutron star-black hole mergers, while LGRBs are associated with the collapse of stripped envelope, massive stars, and the production of energetic supernovae. 

In both short and long GRBs a pair of relativistic, collimated jets are launched from the central engine (i.e., the system composed by a compact object and an accretion disk) and produce the prompt gamma-ray emission. In addition, GRBs are accompanied by a long-lasting afterglow emission (over timescales of $\sim$ yrs in radio bands), produced by the deceleration of the relativistic blast wave. At distances $\gtrsim 10^{16}$~cm, the relativistic jet is decelerated by the large amount of ambient medium piled up by the jet head, and emits a multi-wavelength spectrum extending over most of the electromagnetic spectrum, from X-ray to radio bands \citep[e.g.,][]{SariPiranNarayan_1998, GranotSari_2002, ZhangMacFadyen2009, vanEertenMcfadyen2012, decolle12, decolle12b, Duffell2015, Gill2019}.

The afterglow emission depends strongly on the angular structure of the jet \citep[e.g.,][]{GranotKumar_2003, kumar03, granot2005, salafia2015, Salafiaetal2016}, specially in GRBs seen off-axis as GRB170817A \citep[e.g.,][]{granot2018, Ghirlanda2019, Makhathini2020}, while it is unclear if this is the case also for the prompt emission (see the review by \citealt{salafia2022}).

The structure of the jet launched from the central engine depends on several ingredients such as the presence of a black hole (and its spin) or a neutron star, the structure, mass and angular momentum of the accretion disk, the geometry and intensity of the magnetic field, and the neutrino transport and energy deposition \citep[e.g.,][]{Kathirgamaraju2019, Nathanail2020, Janiuk_2021, gottlieb2022b, gottlieb2022c, gottlieb2022a}.

Several authors have studied the interaction of the jet and the environment (i.e., the progenitor star in LGRBs and the debris of the compact object merger in SGRBs) by employing numerical simulations \citep[e.g.,][]{Morsony2007, Lazzati_2010, Bromberg2011, Mizuta_2013, lopezcamara2013, lopezcamara2016, Hamidani2017,decolle18a, decolle18b, harrison18, lazzati18, Gottlieb2020_hydro_jets,Hamidani2021,Murguia_Berthier_2021,nativi2021interactions, Urrutia2021ShortGRBS,decolle22,HamidaniIoka2022,garciagarcia2022,Suzuki2022}. These works have explored how this interaction depends on the jet properties (luminosity and Lorentz factor) and its magnetization, on the jet opening angle and on the density stratification and velocity of the ambient medium. In all cases, the interaction of the jet with the dense surrounding medium decelerates the jet. The hot shocked material then expands laterally forming a cocoon, which in turn can collimate the jet. Three-dimensional (3D) simulations have shown that the amount of mixing between the jet and the ambient material determines the jet dynamics. Low-density media reduce the jet deceleration and the amount of mixing, allowing the jet channel to remain more stable \citep{duffell18, Hamidani2020, Nathanail2020, Murguia_Berthier_2021, Urrutia2021ShortGRBS, pavan2021, Lazzati_2021_mergerEjecta, GottliebNakarExpandingMedia2021, nativi2021interactions}. 

While the initial jet structure is most likely modified by the interaction with the environment, it remains to be fully understood if different jet structures produced by the central engine lead to different structures at the location where the jet emits electromagnetic radiation. If that is the case, future observations of a large sample of off-axis GRBs, coupled with detailed modeling, will give us insights on the structure of the jet at the launching point (i.e., information on the central engine itself). If not, it will remain impossible to recover this information by electromagnetic signatures alone\footnote{Future gravitational wave detectors as the Big Bang Observer or the DeciHertz Interferometer Gravitational wave Observatory may be able to detect gravitational waves produced directly by the relativistic jets, thus providing us information on the jets while they move inside the dense, optically thick environment (see \citealt{urrutia22}).}.

In the case of SGRBs, \citet{nativi2021interactions} found that  Top-hat or Gaussian jets are indistinguishable after evolving through a high-density medium (the ejecta of a neutron-star merger). \citet{Urrutia2021ShortGRBS}, though, found that while the initial structure of the jets (e.g., Top-hat, Gaussian, and power-law) is not preserved, the jets present different structures even after breaking out of the dense ambient medium.

In addition to the initial conditions of the jet and the environment, three key factors can affect the final structure of the jets. 1) The nature of the energy of the jets, that is, if the jets are pressure dominated (PD) or kinetically dominated (KD) \citep{Marti2016, matsumoto2019}. 2) The magnetization of the jet. \citep{Komissarov2007MagneticAcce,gottlieb2020b_magnetic_jets,Nathanail2020,gottlieb2022c}. 3) The presence of the accompanying type Ic supernova (SN) typically associated to LGRBs \citep[see, e.g.,][]{galama98, hjorth03, malesani04, Pian06}. Although the process leading to the formation of the jet and the SN is not fully understood, they are most likely produced by the same central engine \citep{Woosley1993,MacFadyenWoosley1999}. The interaction between the jet and the SN is not usually considered in numerical simulations, in which it is assumed that the jets propagate through the unperturbed progenitor star. Previous studies have shown that jets propagating through the cavity left from an expanding SN lead to a variety of possible outcomes \citep{Komissarov2007SNjet,decolle22}.

In this paper we study the structure resulting from the propagation of relativistic jets associated to LGRBs with 3D special relativistic hydrodynamical simulations (SRHDs). We run a series of models varying whether the jets are initially PD or KD, the jet structure (Top-hat or Gaussian), and the jet luminosity. We also consider the propagation of jets associated to a SN and show how the jet structure is strongly affected by this interaction. 

The paper is structured as follows: in Section~\ref{sec:methos} we describe the code and the initial conditions employed in our simulations. In Section~\ref{sec:results} we present the results of the numerical simulations. Finally, in Section~\ref{sec:discussion} we discuss the results as well as their implications. In Section~\ref{sec:conclusions} we present the conclusions.

\section{Methods}
\label{sec:methos}
We run a set of 3D numerical simulations by using the adaptive mesh refinement code {\it Mezcal} \citep{decolle12}, which solves the SRHD equations. The SRHD equations are integrated by employing a second-order solver (in space and time). The flux calculation is based on the HLLC Riemman solver \citep{Mignone2005}, which has a low numerical dissipation at the contact discontinuity that allows the development of instabilities during the jet propagation (with respect to more diffusive methods, e.g. the HLL method).

We follow the dynamics of the jet during the first few seconds of propagation through the progenitor star, before the jet breaks out of the stellar surface. The computational domain extends between $x, y \in [-1.5,1.5] \times 10^{10}\,$~cm and $z \in [0,4.5] \times 10^{10}\,$~cm and is resolved in Cartesian coordinates by using $72\times72\times144$ cells along the $x,y,z$ axis at the coarsest level of refinement. The progenitor through which the jet evolves has a radius of $R_{\star}=10^{11}$~cm, and the jet is launched from a spherical boundary (mapped over the Cartesian grid) located at $r_j= 10^9\,$~cm with an opening angle $\theta_j$, being $r=\sqrt{x^2+y^2+z^2}\leq r_j$ and $\theta=\arctan{\left(\sqrt{x^2+y^2}/z^2\right)}\leq \theta_j$. In a cylindrical region, centered at the origin of the coordinate system, extended along the $z$-axis and with a radius of  $5\times 10^{8}$~cm, we set a maximum number of 6 levels of refinement. This corresponds to a minimum cell size $\Delta l \sim 10^7$~cm. Outside this high-resolution region, we set a maximum of 2 levels of refinement, corresponding to $\Delta l\sim 2 \times 10^8$~cm. The refinement criterion is based on density gradients (if $|\nabla \rho |/ \rho \geq 0.4$ the mesh is refined, meanwhile, the mesh is derefined for $|\nabla \rho |/ \rho \leq 0.1$). Reflecting boundary conditions are used at the bottom boundary (i.e., the $z=0$ plane). All other boundaries are set as outflow boundary conditions. The jet propagation is followed during a maximum of 13~s or until the jet arrives to the upper boundary of the computational box. 

The progenitor ($M_{\star}=10 M_\odot$), through which the jet drills, is set by the following density profile \citep{harrison18}
\begin{equation}
    \rho(r) = \frac{A_0}{r^2} \left( 1 - \frac{r}{R_\star} \right)^3,
\end{equation}
where $A_0 = 6 \times 10^{22}$~g cm$^{-1}$ is a normalization constant, and $R_\star=10^{11}$~cm is the stellar radius. The pressure of the progenitor was assumed as $P=10^{-5}\rho(r)c^2$.

For PD jets, we consider the luminosity of the jet expressed as a function of the total energy (kinetic and thermal) per unit volume ($\Delta V$) and unit time ($\Delta t$), i.e.
\begin{equation}
L_j=\left[\rho_j\Gamma_j^2c^2(\Gamma_j-1)+4\Gamma_jP_j\right] \Delta V/ \Delta t\;,
\label{eqn:lum1}
\end{equation}
where we assume an adiabatic index $\Gamma_{\rm ad}=4/3$ for a relativistic gas. The jet luminosity is given by
\begin{equation}
    L_j=\Gamma_\infty \dot{M}c^2\;, 
\label{eqn:lum2}    
\end{equation}
being $\Gamma_\infty$ the asymptotic Lorentz factor and the mass injected in the jet per unit time is
\begin{equation}
    \dot{M}=\left[ \Gamma_j \rho_j \right]\Delta V / \Delta t\;.
\label{eqn:fluxmass1}    
\end{equation}
By substituting expressions \eqref{eqn:lum2} and \eqref{eqn:fluxmass1} in equation \eqref{eqn:lum1}, the pressure of the jet takes the form
\begin{equation}
    P_j \simeq \frac{\rho_j c^2}{4}\left( \frac{\Gamma_\infty}{\Gamma_j} - 1 \right)\;.
    \label{eqn:presurePD}
\end{equation}
Since the unit volume can be expressed as $\Delta V = v_j \Delta t \Delta S $, being $\Delta S$ the surface crossed by the jet and $v_j$ its velocity, we can combine equations \eqref{eqn:lum2} and
 \eqref{eqn:fluxmass1} and obtain the jet density
\begin{equation}
\rho_j = \frac{L_j}{\Gamma_\infty \Gamma_j v_j \Delta S c^2} \;.   
\end{equation}

For KD jets, we assume that the kinetic energy dominates the jet luminosity such that $L_j=\left[ \rho_j\Gamma_j c^2 \left(\Gamma_j -1 \right) \right]v_j\Delta S$. In this case, the jet density is given as
\begin{equation}
    \rho_j = \frac{L_j}{\Gamma_j \left(\Gamma_j -1 \right) c^2 v_j\Delta S } \;.
\end{equation}
In this case, the jet pressure is assumed negligible, i.e. $P_j\sim 10^{-10}\rho_j c^2$. For both PD and KD jets we have $\Delta S = 4\pi (1-\cos\theta_j)r_j^2$.

\begin{figure}
    \centering
    \includegraphics[scale=0.32]{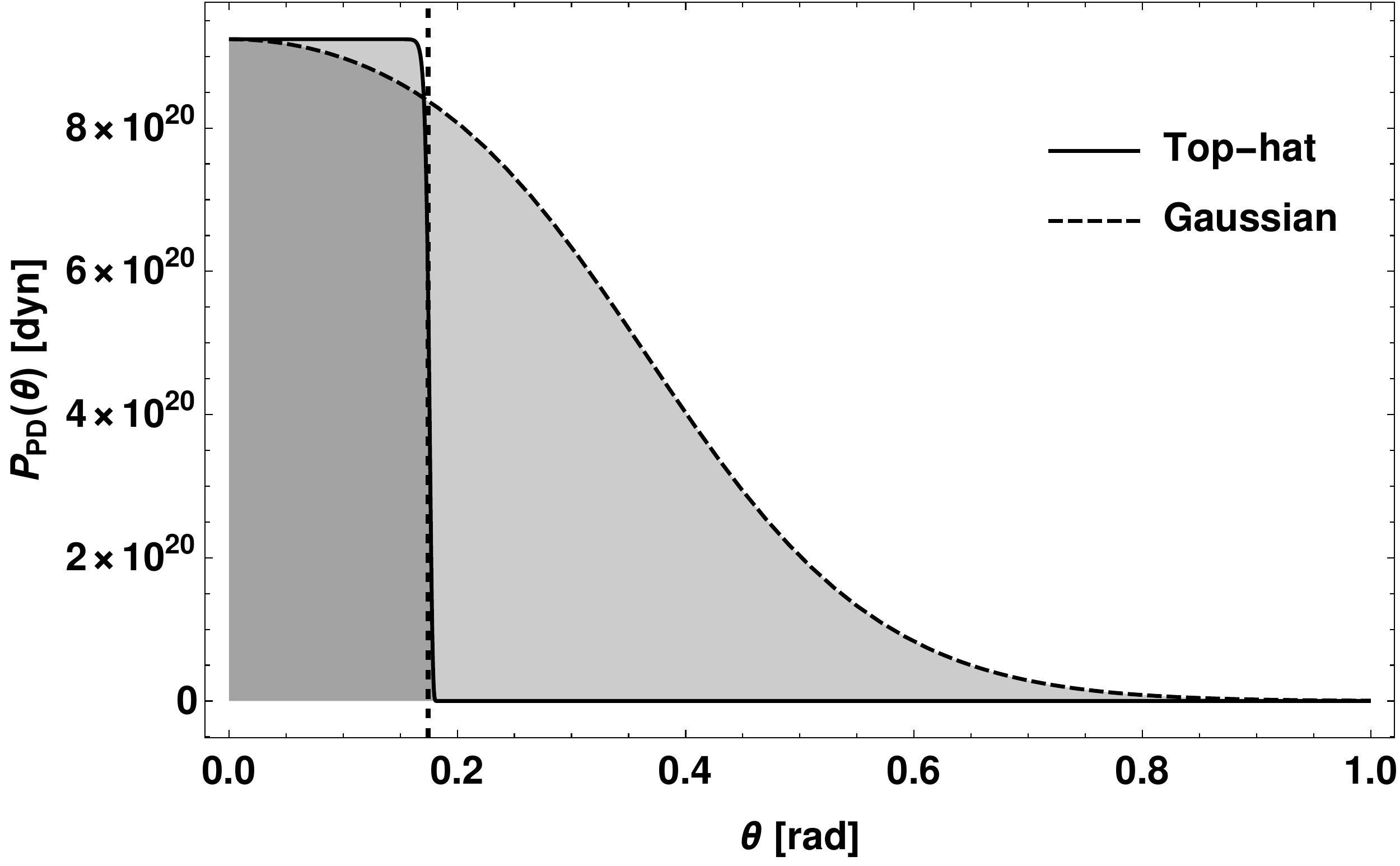}
    \caption{Initial pressure profile for the Top-hat and Gaussian pressure dominated jets ($P_{\rm PD}(\theta)$). The vertical dashed line corresponds to jet core angle $\theta_j=10^\circ$.}
    \label{fig1}
\end{figure}

Also, the jets may either be defined with a Top-hat angular structure, or a Gaussian angular structure. In the Top-hat case the energy, density, and luminosity do not depend on the polar angle (except for the random perturbations described below), i.e. $f(\theta)=1$ for $\theta\le \theta_j$ and $f(\theta)=0$. Meanwhile, in the Gaussian case $f(\theta)=\exp\left[ -0.5(\theta/\theta_j)^2 \right]$. The initial pressure angular profile of Top-hat and Gaussian jets (in the PD models, $P_{\rm PD}(\theta)$) is shown in Figure~\ref{fig1} (the pressure in the KD models
is negligible). For either case, the luminosity and jet Lorentz factors are $L_j=L_{j,0}\,f(\theta)$ and $\Gamma_j=1+(\Gamma_{j,0}-1)f(\theta)$, $\theta_j=10^\circ$, $\Gamma_{j0}=5$, and $\Gamma_{\infty}=400$. The main characteristics of the models\footnote{The jet is launched in the kinetic case with the same Lorentz factor as in the thermal jet (corresponding to $\Gamma=5$), that is, with a Lorentz factor much smaller than what is suggested by observations. We notice that the ram pressure, which depends mainly on the luminosity, regulates the jet dynamics while the jet moves through the dense stellar medium.} are shown in Table~\ref{tab1}.

The jets are launched with either a luminosity $L_{j0}=3.5\times 10^{52}$~erg~s$^{-1}$ (high-luminosity, HL) or $L_{j0}=5.3\times 10^{50}$~erg~s$^{-1}$ (low-luminosity, LL). We also consider the effects that the presence of a supernova has on the jet dynamics. From $r=r_j$, we set a supernova explosion at $t=0\,$s. At $t_{\rm SN}=1\,$s, the jet is injected into the computational box. Following \cite{decolle22}, we inject the SN shock during $t_{\rm SN}=1$~s, with a total energy of   $E_{\rm sn}=4 \times 10^{51}\,$erg and a mass $M_{\rm sn}=0.1M_{\odot}$. We assume that 10\% of the SN energy is thermal, while 90\% is kinetic. The SN is assumed to be asymmetric, with an energy dependence $\propto \cos^2 \theta$. The jet models analysed in this study are shown in Table~\ref{tab1}.

In order to break the axial symmetry we impose small perturbations in the jet launching condition. First, we impose random perturbations of $1\% $ in the jet density such that $\rho'_j=(1+0.01 \Theta)\rho_j$ (being $0\leq\Theta\leq1$ a random number). Second, we impose a small random precession in the jet around the $z$-axis such that the angle between the jet and the $z$-axis is  $\theta_0=\Theta \times 1^\circ$. 
Also, we impose a jet variability luminosity following a Heaviside step function $H(t)$, i.e., $L_{j0}(t)=L_{j0}\,H(t)$ with a period of $0.2$~s as in \cite{lopezcamara2016}.

\begin{table}
\centering
\begin{tabular}{|c||c||c||}
\hline
\textbf{PD or KD} & \textbf{Jet structure} & \textbf{Jet luminosity}  \\
\textbf        &                        &  (erg~s$^{-1}$)  \\ \hline \hline
PD      & Top-hat  & $5.3\times 10^{50}$\\
PD      & Gaussian & $5.3\times 10^{50}$ \\ \hline
PD + SN & Gaussian & $5.3\times 10^{50}$ \\ \hline
KD      & Top-hat  & $5.3\times 10^{50}$ \\
KD      & Gaussian & $5.3\times 10^{50}$ \\ \hline
KD      & Top-hat  & $3.5\times 10^{52}$ \\
KD      & Gaussian & $3.5\times 10^{52}$ \\ \hline\hline
\end{tabular}
\caption{Jet models. Note: the jets with $L=5.3\times 10^{50}$~erg~s$^{-1}$ will be termed as LL while those with $L=3.5\times 10^{52}$~erg~s$^{-1}$ as HL.}
\label{tab1}
\end{table}

\section{Results}
\label{sec:results}

\begin{figure*}
    \centering
    \includegraphics[scale=0.23]{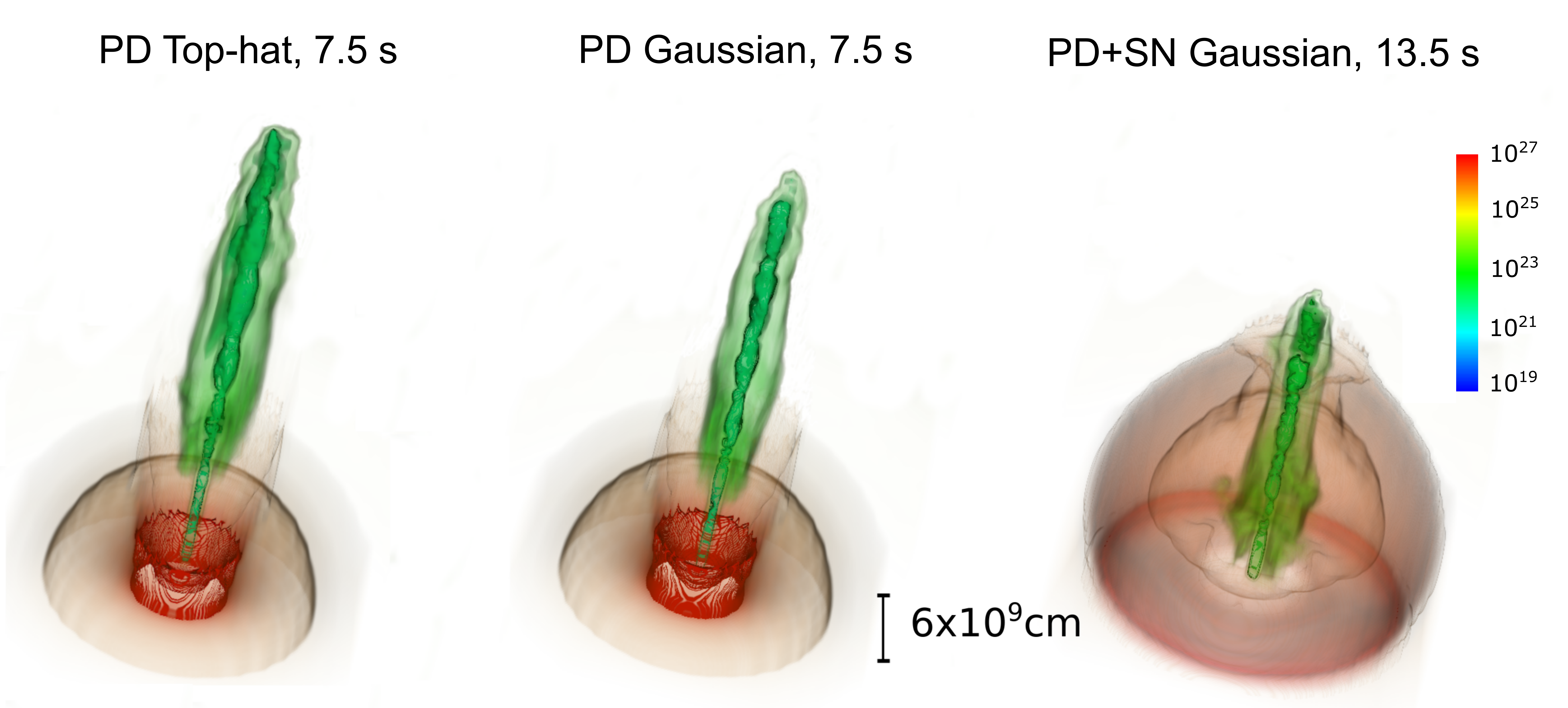}
    \caption{Number density (in units of cm$^{-3}$) volume rendering of the relativistic jets moving through the pre-SN progenitor. The PD Top-hat (left panel) and PD Gaussian (central panel) jet models are shown at $t=7.5\,$s. The case in which a PD Gaussian jet, launched with a delay of 1 s with respect to an associated SN explosion,is shown at 13.5 s (when the jet is about to break-out of the SN shock front, right panel). Note that in all cases the jet is within the progenitor ($R_\star=10^{11}$~cm). }
    \label{fig2}
\end{figure*}

To study the propagation of relativistic jets through the SN progenitor, we first analyzed the evolution of typical jet models employed for long GRBs \citep[e.g.,][]{Morsony2007,Lazzati_2010,Bromberg2011,lopezcamara2013,lopezcamara2016,Hamidani2017,harrison18,Gottlieb2020_hydro_jets,decolle22,Suzuki2022}, this is, a set of models of a PD jet drilling through a massive stellar progenitor (with either a Top-hat or a Gaussian jet structure). Figure~\ref{fig2} shows the evolution of PD jets through the envelope of the progenitor (Top-hat jet: left panel, and Gaussian jet: central panel). Specifically, we show a volume rendering of the number density (cm$^{-3}$) of the relativistic jets moving through the massive stripped-envelope progenitor at $t=7.5$~s. The main components are clearly seen: the progenitor (in brown), the jet and cocoon (in green), and the funnel created by the jet-cocoon as it drills through the progenitor (in red). Note that the jet and cocoon are in all cases within the envelope of the progenitor (which has $R_\star=10^{11}$~cm and for which the breakout time is $\sim$11~s).

\begin{figure*}
    \centering
    \includegraphics[scale=1.2]{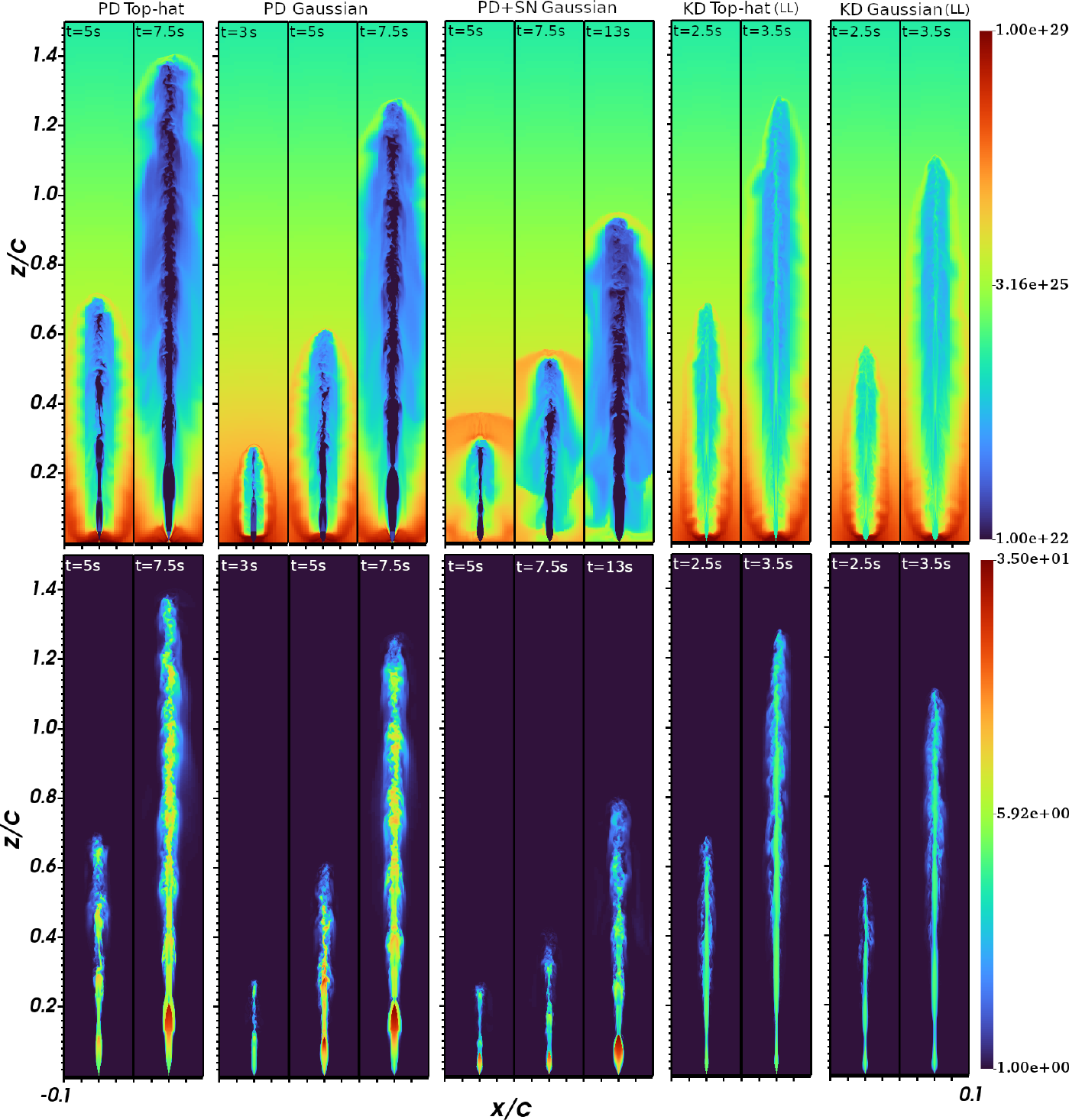}
\caption{Two-dimensional number density (in units of cm$^{-3}$, upper panels) and Lorentz factor (lower panels) stratification maps of the models considered in this study (see Table~\ref{tab1}). From left to right panels: PD Top-hat jet at 5s and 7.5s; PD Gaussian jet at 3s, 5s and 7.5s; PD Gaussian jet with a SN (ejected 1s previously than the jet) at 5s, 7.5s, and 13s; KD Top-hat jet at 2.5s and 3.5s; and KD Gaussian jet at 2.5s and 3.5s. The axis are normalized to c.}
    \label{fig3} 
\end{figure*}

Figure~\ref{fig3} shows 2D number density stratification maps along the $y=0$ cm plane (upper panels), and Lorentz factor stratification maps (lower panels) for different models at various times. For the PD Top-hat and PD Gaussian jets, we show the evolution of a jet (with $L=5.3\times 10^{50}$~erg~s$^{-1}$) through the pre-SN progenitor at 5 s and 7.5 s (Top-hat: left-most panels, and Gaussian: center-left panels). The jet propagation depends on the initial structure (Top-hat vs Gaussian models), being slower by $\lesssim 10$\% the propagation of structured jets with respect to Top-hat jets. While the velocity is slightly smaller in the Gaussian jet, the general morphology remains very similar in the two models. For both the Top-hat and Gaussian models, as the pressure of the jet is larger than the pressure of the cocoon, the jet expands laterally until pressure equilibrium is achieved. Then, the material bounces back towards the axis of symmetry, producing recollimation shocks (visible at the base of the jets). The material in the PD jet is relativistic ($\Gamma \gtrsim 5$), and some acceleration is present beyond the recollimation shocks, with the jet velocity arriving to Lorentz factors of $\sim$ several tens. At larger distances, the jet Lorentz factor drops substantially, going from $\Gamma \sim 30$ below the recollimation shocks, to $\Gamma\sim 5$ at larger values of $z$.

As the PD jet evolves through the envelope, the head of the jet decelerates to sub-relativistic speeds due to the interaction with the dense medium of the progenitor star. The post-shock region is formed by: a) the shocked jet material, formed by material traveling along the jet channel and crossing the reverse shock; and b) the shocked stellar material, formed by stellar plasma heated and accelerated by the forward shock. The post-shock, hot material then expands laterally, forming an extended cocoon which collimates the jet. The cocoon formed by the lateral expansion is sub-relativistic ($\Gamma \gtrsim 1$) as deduced by the fact that the cocoon is not visible in the bottom panels of Figure~\ref{fig3}. Due to the high velocity differences, the contact discontinuity separating the jet and the cocoon is unstable \citep{Matsumoto2021}. 
Then, material close to the contact discontinuity becomes turbulent, which leads to mixing of the stellar and jet material in the cocoon and in the jet. This mixing pollutes the relativistically moving material, dropping substantially its velocity. The latter is visible, for example, at $t=5$ s for the Top-hat PD jet at $z/c\approx 0.5-0.6$. The small perturbations injected in the jet launching boundary lead to a wiggling of the material in the jet which decelerates the jet, and to the formation of blobs along the jet axis.

Since the relativistic jets of long-GRBs may be dominated by kinetic energy, we also analyzed how KD jets evolve through the pre-SN progenitor. Note that the luminosity of the KD jet and the progenitor were the same as those of the PD models. Figure~\ref{fig3} also shows the evolution of KD jets through the progenitor at 5 s and 7.5 s (Top-hat: center-right panels, and Gaussian: right-most panels). Once more, the Gaussian jets are slower than the Top-hat jets, recollimation shocks are present, blobs along the jet axis are formed, and the cocoon is also sub-relativistic. KD jets, though, are quite different both in morphology and dynamics (with respect to the PD jets). KD jets are denser, more collimated, and faster (covering the same distance in about $\sim 0.5$ of the time). The KD jets remain with basically the injected Lorentz factor ($\Gamma \approx 5$) but move faster than the PD jets due to the higher collimation. As an obvious consequence of their low thermal energy, recollimation shocks are much less evident in KD jets. Also, the wiggling of the jet and the creation of blobs are less evident. 

We also explore the (more realistic) scenario in which the GRB and the SN are present. Based on the discussion of the lag between the GRB and the SN of \citet{decolle22}, we assume that the SN takes place $1$~s before the jet. The right panel of Figure~\ref{fig2} shows the evolution of a PD jet. Specifically, we show the volume rendering of the number density (in units of cm$^{-3}$) of a PD Gaussian relativistic jet and its cocoon as it moves through a SN which in turn expands through the progenitor at t = 13.5~s (when it is about to break out of the SN). 

As the SN expands through the progenitor, it produces a dense front which will later affect the evolution of the jet (see Figure~\ref{fig3}). General features of the jets as the collimation, the presence of recollimation shocks, wiggling of the jet and blobs are common also to this model (see the central panels of Figure~\ref{fig3}). On the other hand, the dynamics of jets associated with SNe is dramatically different with respect to the other cases considered. The jet, injected with the same luminosity as the PD model and with a 1~s lag after the SN, first propagates inside the SN cavity which moves through the progenitor. Thus, the jet will be affected by the density stratification produced by the SN. The PD Gaussian jet has to drill through a medium which is $\sim$ one order of magnitude larger when the SN was previously launched compared to when no SN is present (compare the case of the PD+SN Gaussian jet at t = 5 s against the PD Gaussian jet at t = 3 s). At the injection point the density of the star is $\rho_\star(r_j)\approx 37\times 10^{27}$~cm$^{-3}$, dropping to $\approx 8 \times 10^{27}$~cm$^{-3}$ at $r/c \approx 0.07$, then decreasing as $\sim r^{-3}$. On the other hand, the supernova cavity has a density $\rho_{\rm SN} \approx 5\times 10^{27}$~cm$^{-3}$ at $r=r_j$. After $1$~s of evolution, the supernova reverse shock is located at $r/c\approx 0.07$, with a density $\rho \approx 1 \times 10^{27}$~cm$^{-3}$ in the unshocked SN ejecta, and $10-40\times 10^{27}$~cm$^{-3}$ in the shocked ejecta and stellar SN material. That is, the jet moves at the beginning inside a cavity left from the expanding SN, then choke against the expanding SN shock front, which has a much larger density with respect to the progenitor star (at the same radius). 

As a result of the density stratification produced by the SN in its way through the star, the jet propagation is faster as it moves through the SN cavity, and slower once it starts interacting with the denser SN shock front (with respect to the case without SN). Also, as a result of the larger density profile produced by the SN, more energy is deposited in the cocoon (which is slightly more extended). Mixing from the inner part of the cocoon with the jet channel leads to interruption of a continuous supply of relativistic material from the jet into the head, thus leading to a further deceleration of the jet. 

In addition to the models discussed above, we also studied the evolution of a very energetic jet (KD-HL, see Table~\ref{tab1}). This jet has a luminosity much larger than the one expected for GRBs, but will guide our understanding of the dependence of the break out time and the velocity of the jet head as a function of luminosity. In this case, the jet velocity is  close to the speed of light already deep inside the star. The HL jets present basically the same evolution and morphology as the LL case (same density and Lorentz values for the jet and cocoon, and collimation). As expected, the HL jets propagate faster than the LL jets. 

\begin{figure}
    \centering
    \includegraphics[scale=0.35]{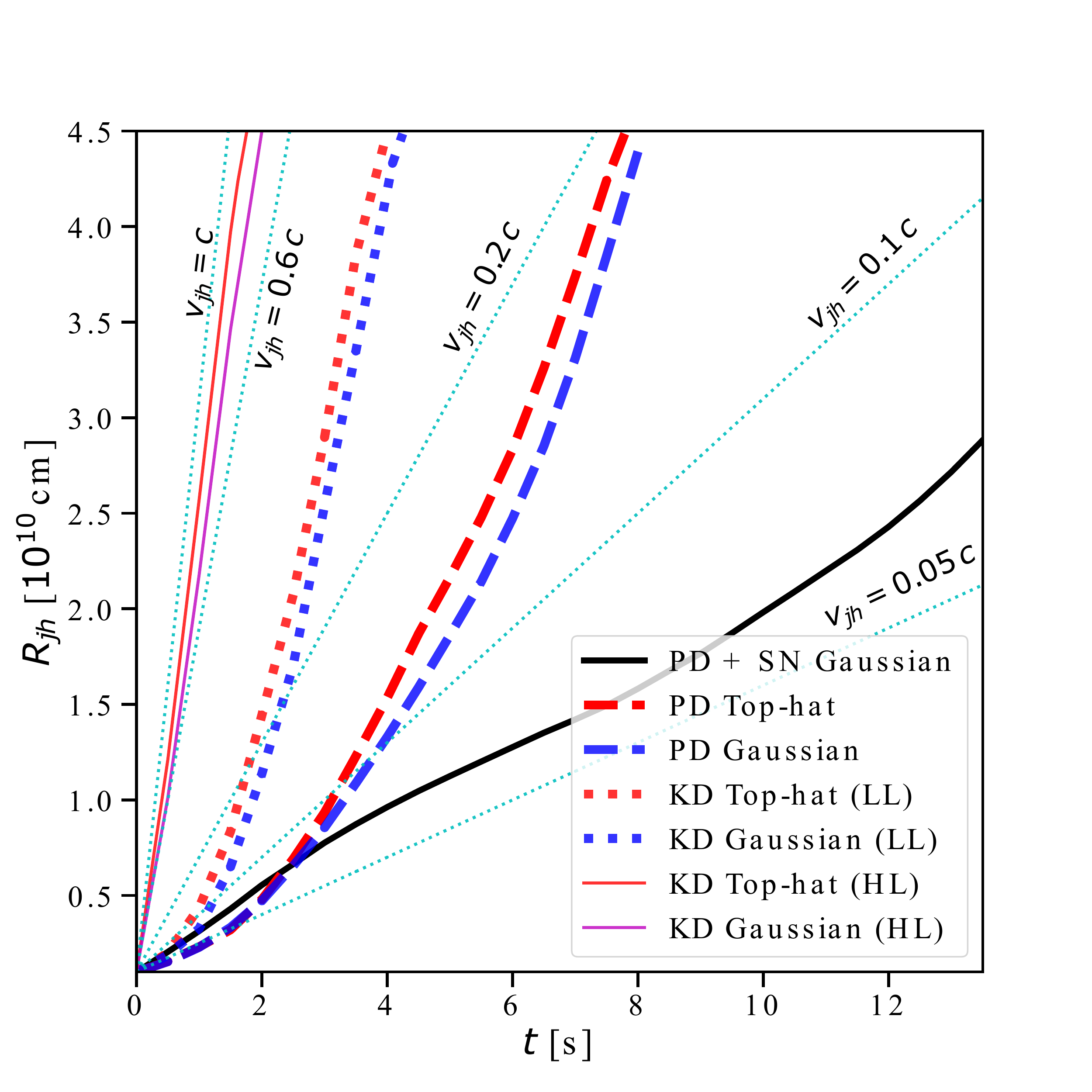}
    \caption{Position of the jet head ($R_{\rm jh}$) as a function of time for the different models. The thin dotted lines correspond to constant head velocities ($v_{\rm jh}$).}
    \label{fig4}
\end{figure}

The most notable effects on the jet head velocity are produced by: a) changing the total jet energy (i.e. more powerful jets move faster); b) if the jet is PD or KD; and c) by the presence of SN moving through the progenitor prior to the jet. To illustrate how the different conditions of the jet (PD or KD, the jet structure, presence of the SN, and the jet luminosity) modify the jet dynamics, in Figure~\ref{fig4} we plot the position of the jet head ($R_{\rm jh}$) along the polar axis as a function of time. In all cases, the KD jets are much faster than the PD jets (with the Top-hat structure producing a faster evolution than a Gaussian structure). For the PD jet models with no SN the $R_{\rm jh}$ accelerate from $v_{\rm jh} \approx 0.05 \;c$ to $v_{\rm jh} \approx 0.6 \;c$ as they approach the stellar surface (notice that these are average velocities, which implies that the local head velocity is larger). The KD jet models accelerate faster and reach higher velocities ($v_{\rm jh} \approx  0.6-0.9 \; c$). The case of a jet propagating into a SN looks quite different. The jet head moves behind the shock front of the SN at a $\sim$ constant velocity of $v_{\rm jh} \lesssim$ 0.1 c. The jet head velocity will remain close to that of the SN until when it breaks out of the SN shock front. Then, it will accelerate as it moves through the much lower density medium.
 
\begin{figure}
  \centering
  \includegraphics[scale=0.292]{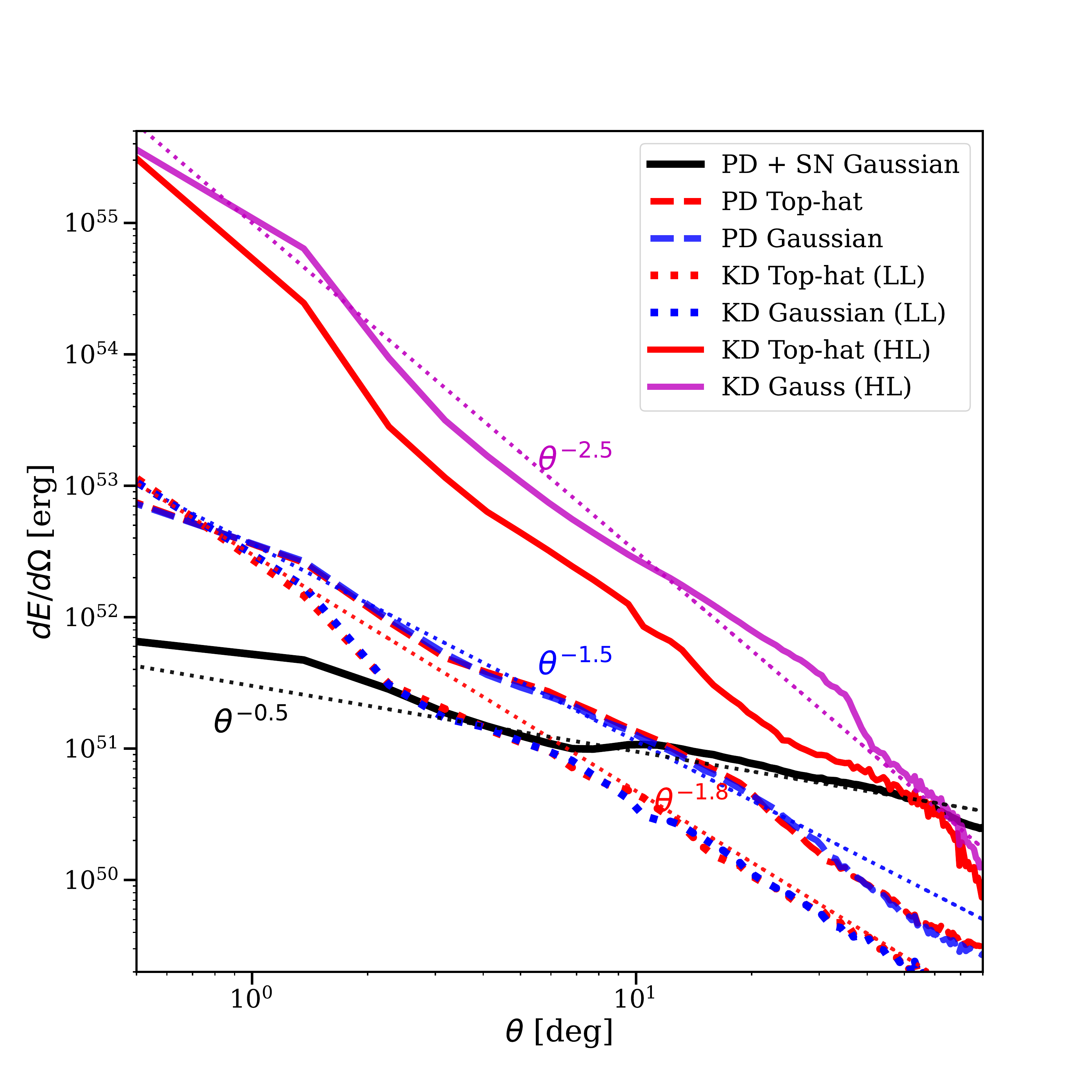} 
  \caption{Energy per solid angle as a function of the polar angle for the models shown in Figure~\ref{fig4} when the correspondent jet head is at $R_{\rm jh}=4.47\times 10^{10}\,$cm. The thin dotted lines correspond to the indicated power laws ($\propto \theta^{-n}$).}
    \label{fig5}
\end{figure}

Figure~\ref{fig5} shows the energy per solid angle computed when the jet head is located at $R_{\rm jh}=4.47\times 10^{10}\,$cm. All curves tend to follow a power-law distribution. The PD jet models follow $dE/d\Omega \propto \theta^{-1.5}$, the KD jet models follow $\theta^{-1.8}$ and $\theta^{-2.5}$ (for the LL and HL, respectively), and the PD+SN jet model follows $\theta^{-0.5}$. Independently if the jets are PD or KD, both Top-hat and Gaussian jets have the same angular distribution. In jets associated to a SN, the energy distribution is much less asymmetric, with the SN dominating at large polar angles.

Although we simulated the propagation of the jet inside the star, we can determine the breakout times by extrapolating the expansion of shock front\footnote{The extrapolation was performed by fitting the curves shown in Figure~\ref{fig4} with the function $R_{\rm jh}(t)= a\, t^b+c\,t$, being $t$ the evolution time. The parameters $a,b,c$ were determined by employing the least square method. The breakout time $t_{\rm bo}$ is then the root of the polynomial equation evaluated at $R_{\rm jh}(t_{\rm bo})=R_{\star}=10^{11}~$cm.}. We found that the Top-hat PD jet has a breakout time $t_{\rm bo}=10.6$~s, similarly to  \cite{harrison18,Morsony2007,lopezcamara2016,Hamidani2017}. We get $t_{\rm bo}=11.2$~s for the gaussian PD models, $t_{\rm bo}=5.8$ s and $t_{\rm bo}=6.1$ s for the Top-hat and Gaussian KD-I models,  $t_{\rm bo}=3.5$ s and $t_{\rm bo}=4.1$ s for the Top-hat and Gaussian KD-II. Finally, the SN model reaches the stellar surface at $t_{\rm bo}=57.3$~s.

\section{Discussion}
\label{sec:discussion}
In this section, we will discuss the differences in the jet dynamics and morphology resulting from the different jet initial conditions (Section~\ref{sec4.1}). Although our simulations show that injecting Gaussian or Top-hat jets lead to a similar jet structure inside the progenitor star, we also discuss under which circumstances different initial jet structures are at least partially preserved once the jet breaks out of the stellar surface and interacts with the circumstellar medium (Section~\ref{sec4.2}). If the initial jet structure is preserved to a certain degree, then, a large sample of observed off-axis GRBs, coupled with numerical calculations, could provide indirect insights on the physics of the central engine and the jet itself.

\subsection{Jet dynamics and morphology}
\label{sec4.1}
We have performed 3D, SRHD simulations of relativistic jets propagating through a massive progenitor. The jets may have initially Top-hat or Gaussian structures, and may be either PD or KD. In addition, the jets may be linked to an accompanying SN and will propagate first through a cavity produced from the SN.

Our simulations show that as the jets evolve through the progenitor, the morphology of the jets are basically independent if their initial structure is Top-hat or Gaussian (see Figure~\ref{fig5}). Gaussian jets though, are slightly slower. On the other hand, whether the jets are PD or KD does produce different morphology and dynamics, being KD jets more collimated and with less energy in the cocoon (see figures~\ref{fig3} and \ref{fig5}).

In Figure \ref{fig5}, the energy distribution for the PD top-hat model was computed at $t\approx 0.75\,t_{\rm bo}$. 
The jet core and power-law distributions observed in our simulations ($1.5-2.5$) are similar to those of (\citealt{Gottlieb2020_hydro_jets}, which find a power-law index of $1-3$).
Based on the temporal evolution shown in Figure~4 of \citet{Gottlieb2020_hydro_jets}, the power-law index is expected to slightly increase as the shock head approaches the surface of the star.
On the other hand, the power-law energy distribution observed in the jet plus supernova case is very different to what is typically associated to isolated jets (as the energy distribution at large angles.

The dynamics of the jet head can be understood by balancing the ram pressure of the material shocked by the forward and reverse shocks, i.e. \citep[see, e.g.,][]{Matzner2003}
\begin{equation}
  \rho_j c^2 h_j \Gamma_j^2 \Gamma_h^2 (\beta_j-\beta_h)^2 + P_j = \rho_a  c^2 h_a \Gamma_h^2 \beta_h^2 + P_a\;,
  \label{eq:vel}
\end{equation}
where $\rho_{j}$, $h_{j}$, $\Gamma_{j}$,  $\beta_{j}$, and $P_{j}$, are the density, enthalpy, Lorentz factor, velocity and pressure of the jet (respectively); $\Gamma_{h}$, $\beta_{h}$ are the Lorentz factor and velocity of the jet head; and $\rho_{a}$, $h_{a}$, and $P_{a}$ are the density, enthalpy, and pressure of the ambient medium. In the case of a strong shock ($P_j \gg P_a$) the ambient mediums satisfies $\rho_a c^2 \gg P_a$ and $h_a\simeq 1$. Also, the jet pressure $P_j$ is a factor $\sim \Gamma_j^2$ smaller than the first term in the left hand side of equation \eqref{eq:vel}. Additionally, if the jet is not cylindrical then its ram pressure at the head location will drops by a geometrical factor $(R_{\rm j}/R_{\rm jh})^2$ (where $R_{\rm j}$ and $R_{\rm jh}$ are the sizes of the jet at its base and at the location of its head, respectively). Thus, taking into account the previous assumptions, the jet velocity will be:
\begin{equation}
   v_h = \frac{v_j}{1+(\rho_a/\rho_j h_j \Gamma_j^2)^{1/2} R_{\rm jh}/r_j}
   \;.
\label{ec: v_j}   
\end{equation}

Equation~\ref{ec: v_j} shows that jets move faster in lower density media (that is, when $\rho_a \lesssim \rho_j h_j \Gamma_j^2$) and when it is more collimated (when $R_{\rm jh}\gtrsim r_j$). Gaussian jets move slower since they have a slightly lower luminosity in the core of the jet, with respect to Top-hat jets. KD jets are faster since they are more collimated than the PD jets (as their lower thermal pressure lead to less prominent recollimation shocks).

\cite{Marti2016} studied the evolution of jets with different types of energy (PD or KD), including also magnetic contributions. In their 2D special relativistic, magneto-hydrodynamic simulations, they found that PD jets have a rich internal structure with several recollimation shocks, while recollimation shocks are mostly suppressed in KD jets. Our 3D SRHD simulations confirm these findings. In addition, we found that KD jets also present mixing in the jet channel (as in the simulations by \citealt{matsumoto2019}), and a larger velocity (by a factor of $\sim 2$) with respect to PD jets.

The mechanism leading to the jet formation is not well established. In the case of magnetohydrodynamic jets, \cite{Komissarov2007MagneticAcce} showed that about $\sim 50$\% of the magnetic energy has been transformed into kinetic energy at distances $r \lesssim 10^9$ cm. In an uncollimated (i.e., conical) PD jet, the acceleration process is very efficient, with a Lorentz factor $\Gamma_j \propto r$. In this case, jet material launched from the central engine with a Lorentz factor of $\sim 10$ close to the central engine ($\sim 10^{6}$ cm) will convert most of its thermal energy into kinetic once it arrives to $\sim 10^7$ cm. If the jets are collimated, the recollimation shocks limit the region where the acceleration happens (see for example Figure~\ref{fig3} in which the jet material is accelerated from $\Gamma_j = 10$ to $\Gamma_j = 30$).

Several studies have discussed the formation and effects of instabilities near the jet core and find that the mixing can strongly decelerate the jet or even stop its propagation  \citep[e.g.,][]{lopezcamara2016,harrison18,matsumoto2019,Gottlieb2020a_intermitent,Gottlieb2020_hydro_jets}. However, our results show that decreasing the thermal pressure in the jet makes it more collimated, stable, and faster during its propagation. Magnetic fields in the jet may also reduce the mixing and stabilize the jet \citep{gottlieb2020b_magnetic_jets,Nathanail2020}. 

The presence of an associated SN strongly affects the propagation of the jet. At the beginning, the jet moves through a cavity (whose extension strongly affects the outcome of the system). When the cavity interacts with the SN shock front, the jet slows down dramatically. It is expected that the jet will break out of the expanding SN, nevertheless, the presence of the SN increases the breakout time and may even inhibit such break out \citep{decolle22}.

\subsection{Is the jet structure determined by the ambient medium  or by the central engine?} 
\label{sec4.2}
In our simulations, different initial structures may lead to different structures only for jets with very large luminosity. This does not imply necessarily that the initial jet structure is completely lost in LGRB jets once the jets break out of the star and accelerate into the low-density, external medium. At least two physical parameters must be considered to properly address this issue: the density of the environment and the jet properties (luminosity, opening angle, Lorentz factor, etc) at the launching point.

If the jet propagates through a low-density ambient medium, then the jet may remain conical during its propagation. Also, the jet head will move with constant speed, the energy deposited into the cocoon will be small, and the jet will remain uncollimated. This corresponds to the case $\tilde{L} > \theta_j^{-4/3}$ (where $\tilde{L} = \rho_j h_j\Gamma_j^2/\rho_a$, see \citealt{Bromberg2011}).
This condition corresponds to $L_j/(c^3 \rho_a S)>\theta_0^{-4/3}$, where $S$ is the surface from where the jet is injected, i.e. $S=4\pi(1-\cos \theta_j) r_j^2 \simeq 2\pi \theta_j^2 r_j^2$. Thus, jets will be uncollimated if 
\begin{equation}
    \rho_a \lesssim 27 {\rm \; g \; cm^{-3}} \left(\frac{L_j}{10^{51} {\rm \; ergs \; s^{-1}}}\right) \left(\frac{\theta_j}{0.1}\right)^{-2/3} \left(\frac{r_j}{10^9 {\rm \; cm}} \right)^{-2}\;.
    \label{eq:rhoa}
\end{equation}
For densities below this value, the jet will be uncollimated, and the initial conditions will be (at least in part) preserved during the jet evolution.
In the stellar progenitor used in this paper, we have $\rho_a = 6\times 10^4$  g cm$^{-3}$ at $r=10^9$~cm. This result is consistent with the fact that the jets in our simulations are collimated. In LGRBs jets, if the jet moves through the dense stellar progenitor, its velocity will be reduced and the structure of the jets at the breakout will be affected by the medium \citep{Irwin2019,Eisenberg2022,gottlieb2022a,Suzuki2022}. 

For SGRBs, equation \ref{eq:rhoa} implies that 
the jets will remain uncollimated if the wind mass-loss $\dot{M}_w$ is:
\begin{equation}
    \dot{M}_w <1.5\times 10^{-3} \left(\frac{L_j}{10^{51} {\rm \; ergs \; s^{-1}}}\right) \left(\frac{\theta_j}{0.1}\right)^{-2/3} \left(\frac{r_j}{10^9 {\rm \; cm}} \right)^{-2} \left(\frac{v_w}{0.3c}\right),
\end{equation}
where we have considered a wind velocity of $0.3 c$.
Envelope masses between $10^{-6}M_{\odot}-10^{-2}M_{\odot}$ are expected in the case of SGRBs \citep{dudi2021,Foucart2021,Dean2021,CombiSiegel2022,MurguiaBerthier2022b,Desai2022,Kullmann2022}. The initial structure of the jet is not affected if it evolves through a thin medium (e.g., $10^{-4}M_{\odot}$ - see \citealt{Urrutia2021ShortGRBS}), while the structure of the jet is affected for denser media ($10^{-2}M_{\odot}$, see \citealt{nativi2021interactions}). 

Since the jet velocity depends non-linearly on the jet luminosity\footnote{From $L_j \propto \rho_j$ and equation~\ref{eq:vel}.}, then the jet dynamics will be very different if the jet is launched with different luminosities or time-luminosity histories \citep[see for example][]{lopezcamara2014, lopezcamara2016}. If the jet luminosity is very large, the jet crosses the inner regions of the stellar envelope at high speeds. While considering a constant, large luminosity is non physical (as the total energy will be much larger than the jet energy inferred from GRB observations, i.e. $10^{51}-10^{52}$ erg), the same outcome can be obtained if the jet luminosity per unit solid angle varies on time (reaching larger initial luminosities then presenting a quick drop). The jet luminosity variability may be due to the varying mass accretion rate onto the compact object \citep[from the depleting accreting disk present around it, see, e.g.,][]{lopezcamara2009}. This kind of time-variable luminosity history has been observed e.g. in relativistic jets produced during some tidal disruption events, with an initial constant luminosity lasting $\sim$week, then dropping as $t^{-5/3}$ (see, e.g., \citealt{decolle20} and references therein). Thus, the larger initial luminosity could help the jet to propagate through the denser inner regions of the progenitor as it deposits less energy in the cocoon, reducing the jet collimation and helping to preserve  the initial structure of the jet.

As shown in Figure~\ref{fig3}, recollimation shocks are present along the jet channel. These recollimation shocks wash out the initial jet structure. Thus, the fate of the structure of the jet is affected by the timescale needed for jet to become uncollimated, which will eventually happen (if the jet lasts long enough) since the expanding cocoon drops its density and pressure with time. On larger timescales, the jet will inject most of its energy into a channel crossing the environment without strongly interacting with it \citep[see, e.g., ][]{Gottlieb2020_hydro_jets}. Most of the energy deposited during this phase will travel unperturbed towards larger distances. Future studies are needed to clarify this issue.

Finally, we notice that GRBs show a large dispersion in the isotropic energy ($10^{48}-10^{54}$ erg, see, e.g., Figure~1 of \citealt{perley14}). Assuming that most of these GRB jets are observed within their core opening angle $\theta_j$, this dispersion is strongly reduced to a jet energy $E_j \sim \theta^2_j E_{\rm iso}/2 \sim 10^{51}$ erg \citep{frail01}. An alternative explanation assumes that GRB jets have a ``universal'' angular structure $dE/d\Omega \propto \theta^{-2}$, and that GRBs with different energies are the produced by jets seen at different observing angles (see, e.g., the review by \citealt{salafia2022} and references therein). From our findings the latter is unlikely, unless the jet structure produced by the central engine and that given by the interaction with the environment are similar for all the GRBs.

\section{Conclusions}
\label{sec:conclusions}
In this work, we study the propagation of 3D SRHD structured LGRB jets evolving through the dense envelope of a progenitor star. The jets were implemented as Top-hat or Gaussian jets whose energy was either PD or KD. In addition, we performed a simulation of a SN plus a lagged jet both propagating through the progenitor.

PD jets present strong recollimation shocks which are absent in KD jets. Then, KD jets drill faster through the progenitor, and are more stable. Also, KD jets are characterized by a lower amount of mixing. While in PD models the jet channel is polluted by cocoon material (diminishing the velocity of the jet), this is not the case in KD jets. Depending on its nature, the jets present different angular energy distribution, being steeper in the KD jet case. Jets propagating firstly through the cavity left from an accompanying SN evolve differently from when no SN is present. While the jet moves faster initially through the SN cavity, it strongly decelerates once it interacts  with the SN shock front. Our simulations illustrate the need to explore further in detail the interaction of the jet with the accompanying SN. 

The initial structure of the jets does not play an important role during their evolution through the progenitor, as the Top-hat and Gaussian jets end very similar to each other. The numerical simulations presented in this work, though, describe the jet dynamics as it moves through the progenitor. The structure of the jet will be preserved if the jet remains uncollimated (i.e., if the jet moves through a low-density environment), which is unlikely in LGRB jets.

As the cocoon expands through the progenitor, the density and pressure of the medium close to the base of the jet drops. Thus, if the jets are long-lasting, they can become uncollimated and preserve the initial conditions at late times. In this case,
off-axis LGRBs with different times duration may show different afterglow behaviors at early times (this is, before the decelerating jet core enters into the observer field of view). The structure of short-lasting LGRBs will reflect the interaction of the jet with the dense medium, while long-lasting LGRBs will depend on the structure of the jet at the injection point. Direct observations of a large sample of GRB afterglows will help to constrain the density stratification of the environment, the initial structure of the jet, and the physics of the central engine itself.

\section*{Acknowledgements}
We thank Ramandeep Gill, Jonathan Granot and Enrique Moreno for useful discussions.
We acknowledge the computing time granted by DGTIC UNAM on the supercomputer Miztli (project LANCAD-UNAM-DGTIC-281), and the support from the UNAM-PAPIIT grants AG100820 and IG100422. 
GU acknowledges support from a CONACyT doctoral scholarship. 
DLC is supported by C\'atedras CONACyT at the Instituto de Astronom\'ia (UNAM).

\section*{Data availability}
The data underlying this article will be shared on reasonable request to the corresponding author.

\bibliographystyle{mnras}
\bibliography{main} 
\bsp	
\label{lastpage}
\end{document}